\documentclass[journal=jpclcd,manuscript=article]{achemso}
\setkeys{acs}{articletitle=true}

\usepackage[version=3]{mhchem} 

\usepackage{amsmath}
\usepackage{amsfonts}
\usepackage{amssymb}
\usepackage{graphicx}
\usepackage{dcolumn}
\usepackage{bm}
\usepackage{subcaption}
\usepackage{algorithm} 
\usepackage{algpseudocode} 
\usepackage{xr}
\usepackage{hyperref}
\usepackage{braket}

\makeatletter
\newcommand*{\addFileDependency}[1]{
  \typeout{(#1)}
  \@addtofilelist{#1}
  \IfFileExists{#1}{}{\typeout{No file #1.}}
}
\makeatother

\usepackage[capitalize]{cleveref}
\crefname{figure}{Fig.}{Figs.}
\Crefname{figure}{Figure}{Figures}
\crefname{table}{Tab.}{Tabs.}
\Crefname{table}{Table}{Tables}
\crefname{equation}{Eq.}{Eqs.}
\Crefname{equation}{Equation}{Equations}
\crefname{section}{Sec.}{Secs.}
\Crefname{section}{Section}{Sections}
\usepackage{xcolor}

\author{Fabijan Pavo\v{s}evi\'{c}}
\email{fpavosevic@gmail.com}
\affiliation{Center for Computational Quantum Physics, Flatiron Institute, 162 5th Ave., New York, 10010  NY,  USA}

\author{Sharon Hammes-Schiffer}
\email{sharon.hammes-schiffer@yale.edu}
\affiliation{Department of Chemistry, Yale University, 225 Prospect Street, New Haven, Connecticut, 06520, USA}

\author{Angel Rubio}
\email{angel.rubio@mpsd.mpg.de}
\affiliation{Max Planck Institute for the Structure and Dynamics of Matter and
Center for Free-Electron Laser Science \& Department of Physics,
Luruper Chaussee 149, 22761 Hamburg, Germany}
\altaffiliation{Nano-Bio Spectroscopy Group and European Theoretical Spectroscopy Facility (ETSF), Universidad del Pa\'is Vasco (UPV/EHU), Av. Tolosa 72, 20018 San Sebastian, Spain}
\alsoaffiliation{Center for Computational Quantum Physics, Flatiron Institute, 162 5th Ave., New York, 10010  NY,  USA}

\author{Johannes Flick}
\email{jflick@flatironinstitute.org}
\affiliation{Center for Computational Quantum Physics, Flatiron Institute, 162 5th Ave., New York, 10010  NY,  USA}

\title[]
  {Cavity-Modulated Proton Transfer Reactions}

\begin{document}



\begin{tocentry}
\begin{figure}[H]
	\begin{center}
		\includegraphics[width=1.7in]{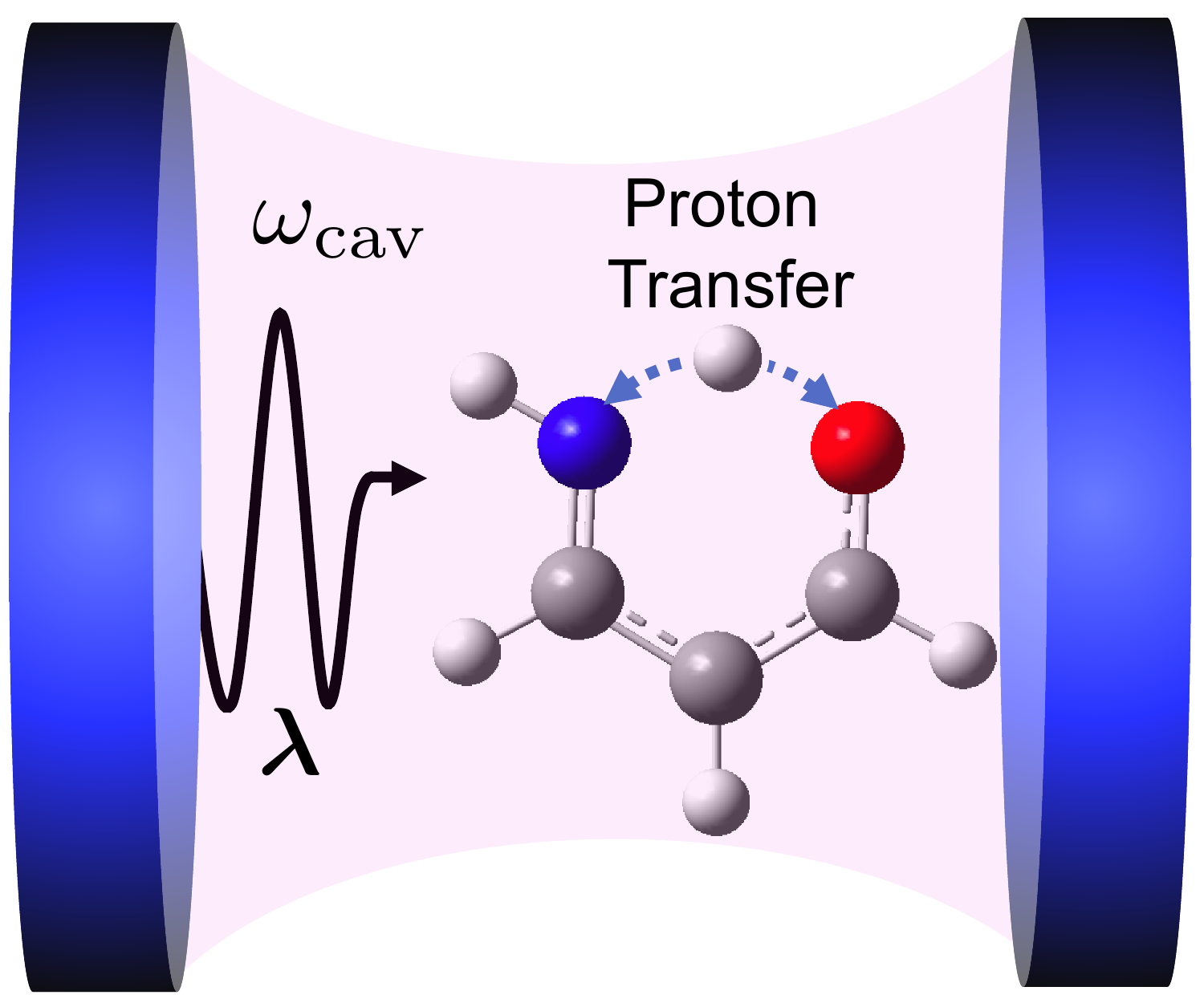}
	\end{center}
\end{figure}
\end{tocentry}

\begin{abstract}

Proton transfer is ubiquitous in many fundamental chemical and biological processes, and the ability to modulate and control the proton transfer rate would have a major impact on numerous quantum technological advances. One possibility to modulate the reaction rate of proton transfer processes is given by exploiting the strong light-matter coupling of chemical systems inside optical or nanoplasmonic cavities. In this work, we investigate the proton transfer reactions in the prototype malonaldehyde and Z-3-amino-propenal (aminopropenal) molecules using different quantum electrodynamics methods, in particular quantum electrodynamics coupled cluster theory (QED-CC) and quantum electrodynamical density functional theory (QEDFT). Depending on the cavity mode polarization direction, we show that the optical cavity can increase the reaction energy barrier by 10--20$\%$ or decrease the reaction barrier by $\sim$5$\%$. By using first principles methods, this work establishes strong light-matter coupling as a viable and practical route to alter and catalyze proton transfer reactions.
\end{abstract}

\maketitle

\section{Introduction}

Proton transfer reactions are essential in many chemical and biological systems due to their key roles in energy conversion processes.\cite{hammes2021virtual} The ability to control the rate of proton transfer reactions can potentially have a large impact on technological advances such as the design of fuel cells, as well as electrochemical and solar energy harvesting devices. For instance, strong light-matter interactions created by quantum fluctuations or external pumping in optical or nanoplasmonic cavities offer a promising route to modulate reaction rates of chemical reactions in a non-intrusive way. Polaritonic chemistry utilizes these strong light-matter coupling effects with chemical systems to catalyze,\cite{lather2019cavity} inhibit,\cite{thomas2016ground} or even modify the overall reaction path \cite{thomas2019tilting} of ground-state and excited-state reactions.~\cite{hutchison2012modifying,Stranius_2018,Munkhbat2018} Besides chemical reactions, strong light-matter coupling with chemical systems has been explored in other context, e.g. intermolecular vibrational energy transfer~\cite{xiang2020intermolecular}, harvesting triplet excitons~\cite{Polak_2020}, self-assembling of microcavities~\cite{Munkhbat2021}, single-molecular strong coupling~\cite{Chikkaraddy2016}, or coupling to nanotips~\cite{Muller2018,roslawska2021mapping}, among others.
These experimental developments have sparked various theoretical developments~\cite{fregoni2018manipulating,lacombe2019exact,climent2019plasmonic,campos2019resonant,li2020resonance,li2021collective,triana2021semi}. The development of predictive theoretical models from first principles is essential for the design and fundamental understanding of chemical processes such as proton transfer inside optical cavities.

One popular method in achieving this goal is the quantum electrodynamics density-functional theory (QEDFT) approach,~\cite{ruggenthaler2014quantum,flick2015kohn,flick2017,ruggenthaler2018quantum,schafer2020relevance} which is a generalization of density functional theory (DFT)~\cite{kohn1999} where both electrons and photons are treated quantum mechanically on the same footing. It is favored for its computational efficiency, which makes it applicable to large system sizes, and has been successfully applied  to a wide range of problems~\cite{flick2018abinito,doi:10.1021/acsphotonics.9b00768} in the regime of strong light-matter interaction. As an extension of electronic DFT, the QEDFT method suffers from similar problems that are inherent to the DFT method, such as challenges associated with dispersion interactions\cite{hermann2017,haugland2021intermolecular} or strongly correlated systems~\cite{burke2012}. In addition, only a small number of approximations for the electron-photon exchange-correlation functional~\cite{pellegrini2015,flick2018abinito,flick2021simple,Schafere2110464118} are currently available for QEDFT. An alternative method to QEDFT is the computationally more expensive, but systematically improvable quantum electrodynamics coupled cluster (QED-CC) method,~\cite{haugland2020coupled,mordovina2020polaritonic} in which the correlation effects between quantum particles (i.e., electrons and photons) are included via the exponentiated excitation cluster operator. The QED-CC method inherits the favorable properties of the electronic CC method, such as the high accuracy, as demonstrated in studies of noncovalent interactions,\cite{haugland2021intermolecular} ionization energies\cite{deprince2021cavity,Pavosevic2021,liebenthal2021equationofmotion}, and excitation energies\cite{Pavosevic2021} in optical cavities, and size-extensivity in molecular systems.\cite{bartlett2007coupled,shavitt2009many,haugland2020coupled} 

In this work, we use the QEDFT formalism and three different QED-CC methods to study the effect of strong-light matter coupling on two different prototypical intramolecular proton transfer reactions in the malonaldehyde and Z-3-amino-propenal (aminopropenal) molecules. To characterize these systems, we compute the reaction energy barrier, which is the energy difference between the transition state and the reactant, as well as the reaction energy, which is the energy difference between the product and the reactant for the asymmetric system. Although hydrogen tunneling is known to play a significant role in these molecular systems, characterizing the potential energy surface in this manner allows a direct comparison between the different methods. Moreover, the reaction energy barrier and other characteristics of the potential energy surface will impact the hydrogen tunneling process. To the best of our knowledge, this work is the first demonstration of proton transfer processes described with quantum electrodynamics $\it ab$ $\it initio$ methods.

\section{Theory}

In general, interactions between molecules and photons inside an optical cavity can be described by the Pauli-Fierz Hamiltonian~\cite{ruggenthaler2014quantum,tokatly2013time,ruggenthaler2018quantum}. We include the optical cavity by coupling the electronic system to a single photon mode.
We further employ the dipole approximation, since the wavelength of the photon mode is much larger than the extent of our molecular system and the length gauge~\cite{craig1998,schafer2020relevance}, where the electric displacement field is coupled to the dipole moment of the system. Additionally we choose the coherent state basis~\cite{haugland2020coupled}. Under these assumptions, the Hamiltonian reads as follows (using atomic units unless otherwise stated)
\begin{equation}
\begin{aligned}
    \label{eqn:PF_Hamiltonian_CSB}
    \hat{H}=h^p_q a_p^q + \frac{1}{2}g^{pq}_{rs} a_{pq}^{rs}+\omega_{\text{cav}}b^{\dagger}b-\sqrt{\frac{\omega_\text{cav}}{2}}(\boldsymbol{\lambda} \cdot \Delta\boldsymbol{d})(b^{\dagger}+b)+\frac{1}{2}(\boldsymbol{\lambda} \cdot \Delta\boldsymbol{d})^2
\end{aligned}
\end{equation}
We note that extensions to multi-mode setups, cavity losses~\cite{wang2021} and correlation effects of nuclei\cite{Hammes-Schiffer19_338,Hammes-Schiffer20_4222} are also possible. The first two terms constitute the electronic Hamiltonian within the Born-Oppenheimer approximation expressed in terms of the second-quantized electronic excitation operators $a_{p_1p_2...p_n}^{q_1q_2...q_n}=a_{q_1}^{\dagger}a_{q_2}^{\dagger}...a_{q_n}^{\dagger}a_{p_n}...a_{p_2}a_{p_1}$ that are defined as a string of fermionic creation and annihilation ($a^{\dagger}$  and  $a$, respectively) operators. Furthermore, $h^p_q=\langle q|\hat{h}^\text{e}|p\rangle$ and $g^{pq}_{rs}=\langle rs|pq\rangle$ denote a matrix element of the core electronic Hamiltonian $\hat{h}^\text{e}$ and a two-electron repulsion tensor element, respectively. The indices $p,q,r,s,...$ denote general electronic spin orbitals, whereas indices $i,j,k,l,...$ and $a,b,c,d,...$  denote occupied and unoccupied electronic spin orbitals, respectively. The third term in this Hamiltonian denotes the photonic Hamiltonian for a single cavity mode with fundamental frequency $\omega_\text{cav}$ expressed in terms of bosonic creation/annihilation ($b^{\dagger}$/$b$) operators. The fourth term describes the dipolar coupling between the electrons and the photonic degrees of freedom. In this term, $\boldsymbol{\lambda}$ is the coupling strength vector that is connected to the field strength of the photon mode~\cite{pellegrini2015,craig1998} and depends, e.g., on the dielectric constant of the material inside the optical cavity and the quantization volume. The dipole fluctuation operator $\Delta\boldsymbol{d}=\boldsymbol{d}-\langle\boldsymbol{d}\rangle$ denotes the change of the dipole operator with respect to its expectation value. The molecular dipole operator $\boldsymbol{d}=\boldsymbol{d}_{\text{e}}+\boldsymbol{d}_{\text{nuc}}$ includes electronic and nuclear components. Finally, the last term in Eq.~\eqref{eqn:PF_Hamiltonian_CSB} describes the dipole self energy arising in the length gauge~\cite{rokaj2018}.

In analogy to conventional electronic structure methods, there are two main ways for solving the Schr\"odinger equation that describes strong light-matter interaction, namely wave function and density functional based formalisms. In the following, we discuss the QED Hartree-Fock and coupled cluster methods, as well as the optimized-effective potential approach (OEP) for QEDFT in more detail.

In the quantum electrodynamics Hartree-Fock (QED-HF) method~\cite{rivera2019,haugland2020coupled}, the wave function ansatz is given as a direct product between an electronic Slater determinant $|0^{\text{e}}\rangle$ and a photon-number state $|0^{\text{ph}}\rangle$ as
\begin{equation}
    \label{eqn:QED-HF}
    |0^{\text{e}}0^{\text{ph}}\rangle=|0^{\text{e}}\rangle\otimes|0^{\text{ph}}\rangle
\end{equation}
where the superscripts $\text{e}$ and $\text{ph}$ denote electrons and photons, respectively. Although this method treats the electrons and photons as uncorrelated particles, it is a useful starting point for correlated methods. Among different approaches, in the QED-CC method~\cite{haugland2020coupled} the correlation effects between quantum particles (electrons and photons) are incorporated via the exponentiated cluster operator 
\begin{equation}
    \label{eqn:QED-T}
    \hat{T}=\sum_{\mu,n}t_{\mu,n}a^{\mu}(b^{\dagger})^n
\end{equation}
that acts on the reference QED-HF wave function as
\begin{equation}
    \label{eqn:QED-CC}
    |\Psi_{\text{QED-CC}}\rangle=e^{\hat{T}}|0^{\text{e}}0^{\text{ph}}\rangle
\end{equation}
In Eq.~\eqref{eqn:QED-T}, the amplitudes $t_{\mu,n}$ are unknown parameters that are determined by solving a set of nonlinear equations~\cite{bartlett2007coupled,shavitt2009many,haugland2020coupled} 
\begin{equation}
    \label{eqn:QED-T-equations}
    \langle0^{\text{e}}0^{\text{ph}}|a_{\mu}(b)^ne^{-\hat{T}}\hat{H}e^{\hat{T}}|0^{\text{e}}0^{\text{ph}}\rangle=\sigma_{\mu,n}
\end{equation}
Moreover, $a^\mu=a_\mu^{\dagger}=\{a_{i}^{a},a_{ij}^{ab},...\}$ is the electronic excitation operator, the index $\mu$ is the electronic excitation rank, and $n$ denotes the number of photons. 

The truncation of the cluster operator at a certain excitation rank $\mu$ and number of photons $n$ establishes the QED-CC hierarchy. Truncation of the cluster operator to include up to single and double electronic excitations along with their interactions with a single photon is expressed as 
\begin{equation}
    \label{eqn:QED-T-21}
    \hat{T}=t^{i,0}_a a_i^a+t^{0,1}b^\dagger+\frac{1}{4}t^{ij,0}_{ab} a_{ij}^{ab}+t^{i,1}_aa_i^ab^\dagger+\frac{1}{4}t^{ij,1}_{ab} a_{ij}^{ab}b^\dagger
\end{equation}
and defines the QED-CCSD-21 method introduced in Ref.~\citenum{haugland2020coupled}. Note that the -$mn$ notation utilized throughout this paper denotes the highest degree of interactions of $m$ electrons with $n$ photons. An extension of the cluster operator defined in Eq.~\eqref{eqn:QED-T-21} to include up to two photons and their interactions with up to two electrons is expressed as
\begin{equation}
\begin{aligned}
    \label{eqn:QED-T-22}
    \hat{T}=t^{i,0}_a a_i^a+t^{0,1}b^\dagger+\frac{1}{4}t^{ij,0}_{ab} a_{ij}^{ab}+t^{i,1}_aa_i^ab^\dagger+\frac{1}{4}t^{ij,1}_{ab} a_{ij}^{ab}b^\dagger+t^{0,2}b^\dagger b^\dagger+t^{i,2}_aa_i^ab^\dagger b^\dagger+\frac{1}{4}t^{ij,2}_{ab} a_{ij}^{ab}b^\dagger b^\dagger
\end{aligned}
\end{equation}
and defines the QED-CCSD-22 method introduced in Ref.~\citenum{Pavosevic2021}. Lastly, truncation of the cluster operator to include interactions between only one electron with up to two photons is expressed as
\begin{equation}
\begin{aligned}
    \label{eqn:QED-T-12}
    \hat{T}=t^{i,0}_a a_i^a+t^{0,1}b^\dagger+\frac{1}{4}t^{ij,0}_{ab} a_{ij}^{ab}+t^{i,1}_aa_i^ab^\dagger+t^{0,2}b^\dagger b^\dagger+t^{i,2}_aa_i^ab^\dagger b^\dagger
\end{aligned}
\end{equation}
and defines the QED-CCSD-12 method first introduced in Ref.~\citenum{white2020coupled} in the context of the description of the electron-phonon interaction. Because the computational cost of the QED-CCSD-$mn$ methods is determined by the number of $t^{ij,n}_{ab}$ amplitude equations that need to be solved, the computational cost of the QED-CCSD-21 and QED-CCSD-22 methods are roughly two and three times higher, respectively, than the computational cost of the QED-CCSD-12 method.

Next, we briefly discuss the optimized-effective potential approach~\cite{pellegrini2015,flick2018abinito} to QEDFT. In contrast to the wave function based methods, such as QED-HF and QED-CC, the QEDFT method obtains solutions to the  Schr\"odinger equation including quantized light-matter interactions in Eq.~\eqref{eqn:PF_Hamiltonian_CSB} in terms of reduced quantities (internal variables). In the length-gauge and dipole approximations, convenient choices for these internal variables are the electron density $n(\textbf{r})$ and the photon displacement coordinate $q=\sqrt{\frac{\hbar}{2\omega_\text{cav}}} \left(b^\dagger + b\right)$~\cite{tokatly2013time,ruggenthaler2014quantum}. Although QEDFT is in principle exact, for practical calculations approximations to the so-called exchange-correlation (xc) potential need to be specified. For QEDFT, these xc potentials must capture not only the correlated nature of the electron-electron interaction as in regular DFT, but also the correlated nature of the quantized electron-photon interaction. So far only a few approximations are available, either in terms of orbital functionals~\cite{pellegrini2015,flick2018abinito} or density functionals~\cite{flick2021simple,Schafere2110464118}. In this work, we choose the optimized-effective potential approximation, which was the first xc potential introduced for QEDFT and is the most established xc potential for problems in QEDFT. This approach is based on the following exchange-correlation energy~\cite{pellegrini2015,flick2018abinito}, which reads for a single photon mode as follows:
\begin{align}
\label{eq:oep}
E_{xc}^{(OEP)} = -\frac{1}{2}\sum_{i,a}|\bra{\varphi_i}\boldsymbol \lambda\cdot \Delta \boldsymbol d \ket{\varphi_a}|^2\left(\frac{\omega_\text{cav}}{\epsilon_a -\epsilon_i + \omega_\text{cav}} -1\right)
\end{align}
where $\epsilon_i$ and $\epsilon_a$ denote occupied and unoccupied Kohn-Sham orbital energies, respectively. We note that  the energy expression in Eq.~\eqref{eq:oep} includes occupied and unoccupied orbitals, but an efficient reformulation in terms of only occupied orbitals is also possible~\cite{flick2018abinito}. This approximation, which is referred to as one-photon OEP, explicitly accounts for one-photon absorption and emission effects and has been shown to be accurate in the weak and strong light-matter coupling regimes.~\cite{pellegrini2015,flick2018abinito,haugland2021intermolecular} For more details on the one-photon OEP approach, we refer the reader to Refs. \citenum{pellegrini2015,flick2018abinito,octopus3,haugland2021intermolecular,schafer2020interface}.

\section{Results and Discussion}
The QED-CCSD-$mn$ methods have been implemented in an in-house developmental version of the Psi4NumPy quantum chemistry software~\cite{smith2018psi4numpy}. The implemented QED-CCSD-$mn$ methods, along with the QEDFT method, were used to calculate the reaction energy diagrams for proton transfer in malonaldehyde and aminopropenal. All calculations were performed on the geometries optimized at the conventional electronic CCSD/cc-pVDZ\cite{dunning1989gaussian} level using the Gaussian quantum chemistry software.\cite{G16A03} The geometry optimizations were performed using the standard optimization procedures\cite{schlegel1982optimization,peng1993combining} with default parameters as implemented in Gaussian 16.\cite{G16A03} The characters of the stationary structures on the potential energy surface (i.e., reactants, transition states, and products) were  confirmed by performing the harmonic frequency analysis. We note however, that the stationary points can be different between HF, CCSD, and DFT. The QED-HF and the QED-CCSD-$mn$ calculations were performed by employing the cc-pVDZ basis set.\cite{dunning1989gaussian} The QEDFT calculations were performed using the Octopus code~\cite{octopus3} with the single-photon OEP implementation described in Refs. \citenum{flick2018abinito,octopus3}. This formulation describes both the electron-electron interaction and the electron-photon interaction consistently within the OEP approach. If no electron-photon coupling is present, this approach reduces to the electronic OEP framework~\cite{kuemmel2008}.
In all calculations with the Octopus code, we used a real-space grid with spheres of 6~\AA~around each atom and a grid spacing of 0.15~\AA~, as well as Troullier-Martins pseudopotentials~\cite{troullier1993} to describe the core electrons.

\begin{figure*}[ht!]
  \centering
  \includegraphics[width=6.5in]{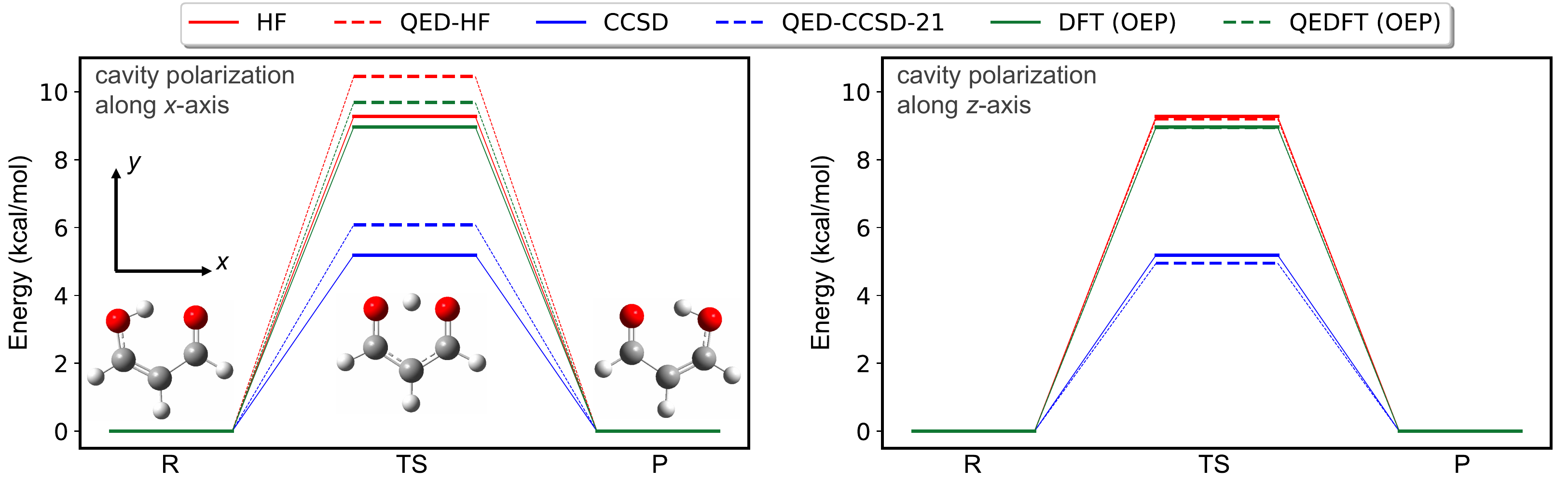}
  \caption{Reaction diagram for proton transfer in  malonaldehyde calculated with different levels of theory outside (solid) and inside (dashed) an optical cavity. The calculations employ the cavity parameters $\omega_{\text{cav}}=3$ eV and $\lambda=0.1$ a.u. with the photon mode polarized in the $x$ (in the molecular plane and parallel to the proton transfer direction) (left), and $z$ (perpendicular to the molecular plane) (right) directions. The left panel contains the images of the reactant (R), transition state (TS), and product (P) structures along with the coordinate frame. }
  \label{fig:malonaldehyde}
\end{figure*}

Figure \ref{fig:malonaldehyde} depicts the reaction energy diagram for the proton transfer process in malonaldehyde calculated with the HF, CCSD, and DFT methods (solid lines) as well as with their QED counterparts (dashed lines). We only include QED-CCSD-21 results in this figure for clarity, and the effect of the cavity calculated with the QED-CCSD-12 and QED-CCSD-22 methods is given in Table \ref{table:reaction_barrier_malonaldehyde}. The QED calculations were performed in an optical cavity with frequency $\omega_\text{cav}=3$ eV and coupling strength $\lambda=0.1$ a.u. with the light polarized along the $x$ direction (in the molecular plane and parallel to the direction of proton transfer) and $z$ direction (perpendicular to the molecular plane). The result with the cavity mode polarized in the $y$ direction is qualitatively similar to the result with the cavity mode with polarization along the $x$ direction and is therefore not included in the figure, but it is provided in Table \ref{table:reaction_barrier_malonaldehyde}. The reaction barrier for this reaction strongly depends on the choice of electronic structure method, where both the HF and DFT methods overestimate this barrier, and the CCSD provides a barrier that is in good agreement with the barrier calculated with the coupled cluster singles and doubles with perturbative triples (CCSD(T)) method.\cite{mil2008tunneling} Since we calculated the energies of the transition state with the CCSD geometry, we can note that in case of HF the overestimation of the barrier is mainly due to the missing correlation effects since a transition state at the HF level changes the barrier compared to the CCSD geometry by $\sim$1 kcal/mol. All of the employed QED methods predict the same trend: the reaction barrier increases if the cavity mode is polarized in the molecular plane ($x$ and $y$ directions), and the reaction barrier decreases if the cavity mode is polarized perpendicular to the molecular plane ($z$ direction). 

\begin{table}[ht]
\centering
\begin{tabular}{c  c  c  c}
\hline
method & $x$ direction & $y$ direction & $z$ direction\\
\hline\hline
QED-HF      & 1.18 & 0.15 & -0.07\\
QED-CCSD-12 & 1.01 & 0.12 & -0.24\\
QED-CCSD-21 & 0.89 & 0.05 & -0.24\\
QED-CCSD-22 & 0.84 & 0.04 & -0.24\\
QEDFT (OEP)      & 0.72 & 0.01 & -0.01\\
\hline
\end{tabular}
\caption{Change in the Reaction Energy Barrier (TS) for Proton Transfer in Malonaldehyde inside an Optical Cavity.\textsuperscript{a}}
\textsuperscript{a}\small Relative energies are calculated as the difference between the reaction barrier obtained with the QED method and the corresponding conventional electronic structure method. Relative energies are given in kcal/mol.\\
\label{table:reaction_barrier_malonaldehyde}
\end{table}

As shown in Table \ref{table:reaction_barrier_malonaldehyde}, the greatest effect of the cavity mode (quantum fluctuations) on the reaction energy barrier is observed with the cavity mode polarized along the $x$ direction, followed by the $y$ and $z$ directions. This observation is in agreement with previous findings that molecules with the largest fluctuations of the dipole self energy, $\frac{1}{2}(\boldsymbol{\lambda} \cdot \Delta\boldsymbol{d})^2$, will experience the greatest effect of the cavity.\cite{deprince2021cavity} The changes of this quantity between the transition state structure and the reactant structure calculated with the QED-HF method for cavity modes polarized along the $x$, $y$, and $z$ directions are 1.06 kcal/mol, 0.11 kcal/mol, and -0.07 kcal/mol, respectively, and therefore account for most of the overall changes seen in Table \ref{table:reaction_barrier_malonaldehyde}. Recently, the connection between the one-photon OEP energy and the dynamical polarizabilities has been worked out~\cite{flick2021simple}. Thus, we compare the static polarizabilities and find the calculated absolute values of the static polarizabilities within the QED-HF method are ordered as $\alpha_{xx} > \alpha_{yy} > \alpha_{zz}$ in agreement with the reasoning on the dipole self energy. In the case of the in-plane polarized cavity mode, the inclusion of correlation effects between the quantum particles with either the QED-CCSD-$mn$ or QEDFT methods reduces the effect of the cavity on the reaction energy barrier. Moreover, inclusion of correlation effects with the QED-CCSD-$mn$ methods stabilizes the transition state for the cavity mode with polarization along the z direction and further reduces the reaction barrier compared to the QED-HF method.

Table \ref{table:reaction_barrier_malonaldehyde} also shows that for the specified cavity parameters and in-plane polarized light, the QED-CCSD-21 method performs more similarly to the QED-CCSD-22 method than to the QED-CCSD-12 method. Deeper analysis indicates that the key to this similarity is the $t^{ij,1}_{ab}$ amplitudes that describe correlation effects between electronic double excitations and single photon processes, which are not present in the QED-CCSD-12 method. All three of the QED-CCSD-$mn$ methods predict the same effect of the cavity mode with polarization along the $z$ direction. The QEDFT method with the OEP functional predicts the smallest effect of the cavity among the investigated methods.

\begin{figure}[ht]
  \centering
  \includegraphics[width=3.25in]{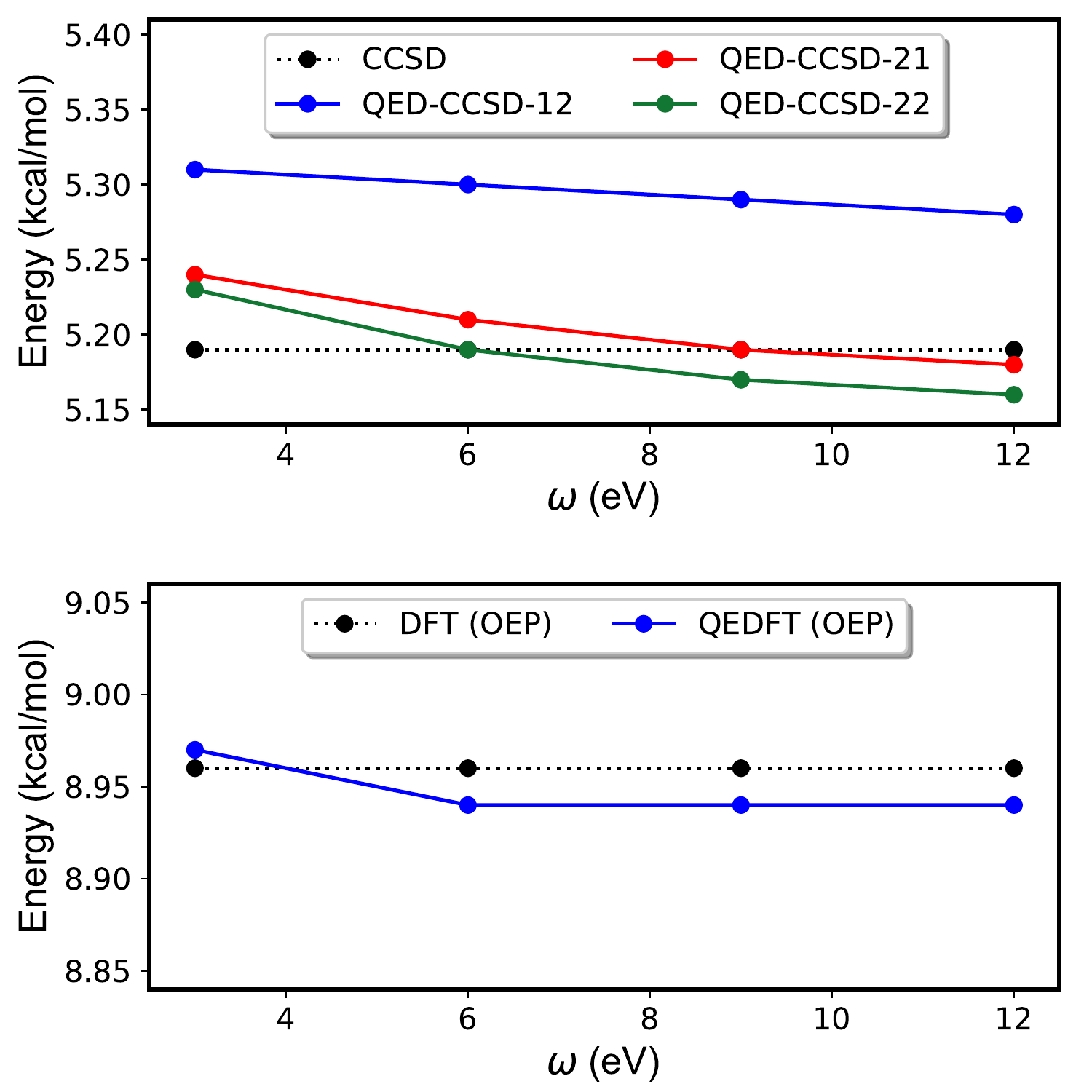}
  \caption{Proton transfer reaction barrier for malonaldehyde as a function of cavity frequency. Dotted lines correspond to calculations outside the cavity, whereas solid lines correspond to calculations inside the cavity with the cavity mode polarized along the $y$ direction and with coupling strength 0.1 a.u. Reaction barriers calculated with the conventional CCSD and DFT methods, along with their QED counterparts, are given in the upper and lower panels, respectively.}
  \label{fig:barrier_vs_omega}
\end{figure}

Figure \ref{fig:barrier_vs_omega} shows the change in the reaction barrier for malonaldehyde calculated with the CCSD methods (upper panel) and the DFT methods (lower panel) as the cavity frequency is increased from 3 eV to 12 eV by 3 eV increments. The QED barriers (solid lines) were calculated in a cavity with the mode polarized along the $y$ direction with coupling strength 0.1 a.u. The reaction barrier energies calculated with the conventional CCSD and DFT methods are depicted with the dotted black line and are independent of the cavity frequency. All of the QED methods except for the QED-CCSD-12 method predict that the barrier will change character  from being higher than to being lower than the energy barrier outside the cavity within this range of cavity frequencies. This change in character occurs at 4 eV, 6 eV, and 9 eV for QEDFT, QED-CCSD-22, and QED-CCSD-21, respectively. Therefore, the QED-CCSD-12 model does not seem to be accurate enough to capture the subtle effects caused by the optical cavity. We have also calculated the change in the reaction barrier for the different QED-CCSD-$mn$ methods and the QEDFT method as a function of the coupling strength, with results given in Fig. S1. This figure shows that the reaction barrier can increase by $\sim$3 kcal/mol for QED-CCSD-$mn$ and by $\sim$4 kcal/mol for QEDFT in the case of very large values of coupling strength. We find a quadratic scaling with lambda for the QEDFT-OEP results, which can be expected from Eq.~\eqref{eq:oep}. Moreover, we also find that the QED-CCSD-21 method is in good agreement with the QED-CCSD-22 method for smaller values of the coupling strength ($<$0.1), whereas for larger values of the coupling strength ($>$0.15), the QED-CCSD-21 method approaches the less accurate QED-CCSD-12 method. Thus, the QED-CCSD-21 method should be used with caution when the light-matter coupling is strong.

\begin{figure*}[ht!]
  \centering
  \includegraphics[width=6.5in]{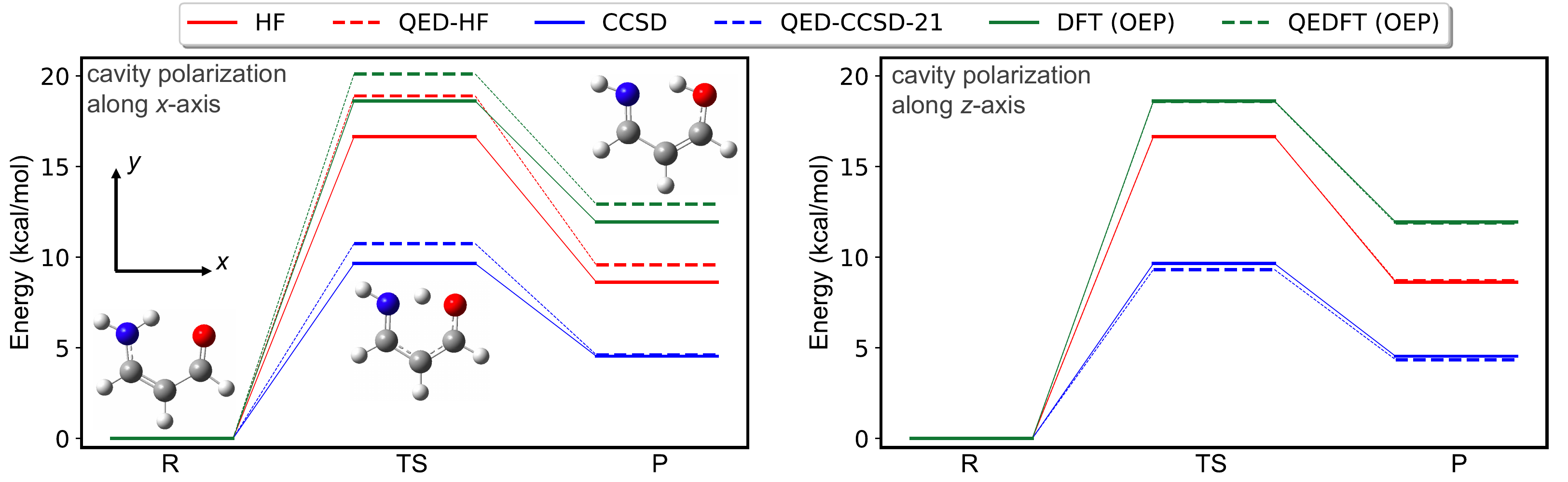}
  \caption{Reaction diagram for proton transfer in aminopropenal calculated with different levels of theory outside (solid) and inside (dashed) an optical cavity. The calculations employ the cavity parameters $\omega_{\text{cav}}=3$ eV and $\lambda=0.1$ a.u. with the photon mode polarized in the $x$ (in the molecular plane) (left), and $z$ (perpendicular to the molecular plane) (right) directions. The left panel contains the images of the reactant (R), transition state (TS), and product (P) structures along with the coordinate frame.}
  \label{fig:aminopropenal}
\end{figure*}

As another example, we investigate the cavity effect on proton transfer in aminopropenal. Unlike malonaldehyde, aminopropenal is asymmetric and has different reactant and product structures, as shown in Fig. \ref{fig:aminopropenal}. Note that the optimized geometry of the reactant shown in Fig. \ref{fig:aminopropenal} was found to have the lowest energy of all considered stationary points. This figure depicts the reaction energy diagram for proton transfer in aminopropenal calculated with the HF, CCSD, and DFT methods outside (solid lines) and inside (dashed lines) the cavity with the mode polarized along the $x$ and $z$ directions, and with frequency $\omega_\text{cav}=3$ eV and coupling strength $\lambda=0.1$ a.u. The changes in reaction energies and barriers for the cavity mode polarized along the $x$, $y$, and $z$ directions calculated with all of the QED methods investigated are provided in Table \ref{table:reaction_barrier_aminopropenal}. As in the case of malonaldehyde, all of these QED methods predict the same trend for the change in the reaction barrier. In particular, the reaction barrier increases if the cavity mode is polarized in the molecular plane ($x$ and $y$ directions), and the reaction barrier decreases if the cavity mode is polarized  in perpendicular to the molecular plane ($z$ direction). Again this trend is consistent with the change of the $\frac{1}{2}(\boldsymbol{\lambda} \cdot \Delta\boldsymbol{d})^2$ term between the transition state and the reactant. For example, the changes of this term between the transition state and the reactant calculated with the QED-HF method are 1.88 kcal/mol, 0.60 kcal/mol, and -0.01 kcal/mol for a  cavity mode with polarization along the $x$, $y$, and $z$ directions, respectively. Moreover, as for malonaldehyde, the HF and DFT methods overestimate the reaction barrier,  predicting it to be roughly two times larger than the barrier computed with the CCSD method.

\begin{table}[htbp]
\centering
\begin{tabular}{c c c c c c c}
\hline
 method & \multicolumn{2}{c}{$x$ direction} & \multicolumn{2}{c}{$y$ direction} & \multicolumn{2}{c}{$z$ direction} \\
 \hline\hline
 &      TS    &     $\Delta E$     &     TS      &     $\Delta E$     &     TS      &      $\Delta E$    \\
  \hline
QED-HF &     2.00      &      0.94    &     0.59      &   0.37       &     -0.01      &     0.06     \\
QED-CCSD-12 &     1.12      &     -0.07     &     0.09      &       -0.34   &      -0.33     &    -0.19    \\
QED-CCSD-21 &     1.10      &     0.08     &     0.06      &     -0.25     &      -0.33     &     -0.20     \\
QED-CCSD-22 &     1.07      &     0.13     &     0.06      &      -0.22    &      -0.33     &     -0.20     \\
QEDFT (OEP) &     1.49      &     0.97     &     0.13      &    -0.07      &      -0.04     & -0.07\\
\hline
\end{tabular}
\caption{Change in the Reaction Energy Barrier (TS)\textsuperscript{a} and Reaction Energy ($\Delta E$)\textsuperscript{b} for Proton Transfer in Aminopropenal inside an Optical Cavity.\textsuperscript{c}}

\textsuperscript{a}\small Relative energies are calculated as the difference between the reaction barrier obtained with the QED method and the corresponding conventional electronic structure method. \\

\textsuperscript{b}\small Relative energies are calculated as the difference between the reaction energy (i.e., the difference between the energies of the product and reactant) obtained with the QED method and the corresponding conventional electronic structure method.\\

\textsuperscript{c}\small Relative energies are given in kcal/mol.\\

\label{table:reaction_barrier_aminopropenal}
\end{table}

The greatest increase in the reaction barrier for the cavity mode with polarization along the $x$ direction is calculated with the QED-HF method. Inclusion of the single-photon effects with the QEDFT-OEP method decreases this value by 0.51 kcal/mol, whereas the  QED-CCSD-$mn$ methods reduce this change by $\sim$1 kcal/mol. Similarly, for the cavity mode with polarization along the $y$ direction, the QED-HF method shows the greatest increase in reaction barrier due to the presence of the cavity, whereas inclusion of the correlation effects between quantum particles at the QEDFT and QED-CCSD-$mn$ levels increases the barrier by a smaller amount. For the cavity mode with polarization along the $z$ direction, the QED-HF and QEDFT methods show a very small decrease of the reaction barrier, whereas this change is more pronounced for the QED-CCSD-$mn$ methods and remains the same within 0.01 kcal/mol for all three definitions of the cluster operator. 

Next, we discuss the cavity effect on the reaction energy for the proton transfer reaction in aminopropenal. The reaction energy is calculated as the energy difference between the product and the reactant. Unlike in the case for the changes in the energy barriers, Fig. \ref{fig:aminopropenal} and Table \ref{table:reaction_barrier_aminopropenal} indicate that the calculated change in reaction energy in the presence of the optical cavity calculated with different QED methods do not follow the same trend. In the case of the cavity mode with polarization along the $x$ direction, both the QEDFT and QED-HF methods predict that the cavity significantly increases the reaction energy, whereas the QED-CCSD-$mn$ methods predict a much less pronounced change. Additionally, the QED-CCSD-12 method even predicts a slight decrease in the reaction energy. In the case of the cavity mode polarized along the $y$ and $z$ directions, all of the correlated methods predict a decrease of the reaction energy in the presence of the optical cavity, whereas the QED-HF method predicts the opposite trend.

Lastly, we calculate the Rabi splitting for malonaldehyde and  aminopropenal by employing the linear response QEDFT method~\cite{doi:10.1021/acsphotonics.9b00768}. In this calculation, we tune the cavity mode into resonance with their respective HOMO-LUMO excitation energies and vary the electron-photon coupling strength $\lambda$. The computational details are provided in the SI. From the results given in Fig. S2, we find that for the coupling strength of $\lambda=0.1$ a.u., the ratio of the Rabi splitting to the cavity frequency is $\sim$11\%, which is within the range of experimentally observed values in the single-molecule~\cite{Chikkaraddy2016} and collective strong coupling \cite{hutchison2012modifying} limits. Therefore, the choice of the coupling strength of $\lambda=0.1$ a.u. that has been applied to the ground-state calculations throughout this paper is a reasonable choice.

\section{Conclusion}
In this work, we study the effect of an optical cavity on proton transfer in malonaldehyde and aminopropenal with different QED $\it ab$ $\it initio$ methods. In particular, herein we implement and test the QED-HF, QED-CCSD-12, QED-CCSD-21, and QED-CCSD-22 methods. We show that the optical cavity with the mode polarized in the molecular plane can increase the reaction energy barrier by 10--20$\%$, whereas polarization of the cavity mode perpendicular to the molecular plane can decrease the reaction barrier by $\sim$5$\%$. All of the calculations indicate that the QEDFT and QED-CCSD-$mn$ methods show the same qualitative effect on the reaction barrier and the reaction energy due to the optical cavity with the exception of the QED-CCSD-12 method, which includes interactions of only one electron with the photon(s). This finding indicates that the correlation effects between the quantum particles play an important role for the accurate prediction of the changes in the presence of an optical cavity and that a significant amount of correlation energy must be recovered for which interactions between two electrons and the photon(s) are essential. Therefore, the QED-HF and QED-CCSD-12 methods are not recommended for these types of systems. The developed methods also set the stage for the understanding of strong light-matter coupling in solid state systems and materials including the effects of losses of the cavity and materials. The work presented herein will serve as the basis for future experimental and theoretical investigations of proton transfer reactions inside an optical cavity and as a benchmark resource for the QEDFT and QED-CCSD methods.

\begin{acknowledgement}
We acknowledge financial support from the European Research Council (ERC-2015-AdG-694097), Grupos Consolidados (IT1249-19), the Cluster of Excellence 'CUI: Advanced Imaging of Matter' of the Deutsche Forschungsgemeinschaft (DFG) - EXC 2056 - project ID 390715994, and the U.S. Air Force Office of Scientific Research under AFOSR Award No. FA9550-18-1-0134 (S.H.-S.).  We also acknowledge support from the Max Planck–New York Center for Non-Equilibrium Quantum Phenomena. 
The Flatiron Institute is a division of the Simons Foundation.
\end{acknowledgement}

\noindent\textbf{Supporting Information Available}: The supporting information includes: energy barrier dependence on the coupling strength; Rabi-splitting for different electron-photon coupling strengths.

\noindent\textbf{Keywords}: ab initio calculations, optical and nanoplasmonic cavity, proton transport, quantum chemistry, strong light-matter interaction.

\noindent\textbf{Conflict of interest}\\
The authors declare no conflict of interest.

\linespread{1}\selectfont
\bibliography{Journal_Short_Name,references}{}

\end{document}